\documentclass[amssymb,aps,twocolumn,showpacs,superscriptaddress,nofootinbib]{revtex4}
\usepackage[]{graphicx}
\usepackage[]{amsmath}
\usepackage{hyperref}

\def\ket#1{| #1 \rangle}

\def\bracket#1#2{\langle #1 | #2 \rangle}
\def\kb#1#2{| #1 \rangle\!\langle #2 |}

\def\cO{\mathcal{O}}

\def\cZ{\mathcal{Z}}

\def\Tr{\mathrm{Tr}}

\newcommand{\C}{\mathbb{C}}
\newcommand{\<}{\langle}
\renewcommand{\>}{\rangle}
\newcommand{\Log}{\mathrm{Log}}
\newcommand{\tr}{\mathrm{Tr}}

\begin{document}

\title{Sampling from the thermal quantum Gibbs state \\ and evaluating partition functions with a quantum computer}
\author{David Poulin}
\affiliation{D\'epartement de Physique, Universit\'e de Sherbrooke, Qu\'ebec, Canada}
\author{Pawel Wocjan}
\affiliation{School of Electrical Engineering and Computer Science, University of Central Florida, FL, USA}

\date{\today}

\begin{abstract}
We present a quantum algorithm to prepare the thermal Gibbs state of interacting quantum systems. This algorithm sets a universal upper bound $D^\alpha$ on the thermalization time of a quantum system, where $D$ is the system's Hilbert space dimension and $\alpha \leq \frac 12$ is proportional to the Helmholtz free energy density of the system. We also derive an algorithm to evaluate the partition function of a quantum system in a time proportional to the system's thermalization time and inversely proportional to the targeted accuracy squared. 
\end{abstract}


\maketitle

The partition function at inverse temperature $\beta$ is of central interest in statistical physics because all thermodynamic variables such as energy, heat capacity, entropy, and correlation functions can be derived from it. For interacting systems, the partition function involves a sum over an exponential number of terms, so it cannot be computed exactly except in few special cases. Thus, approximation schemes are required to estimate these physical quantities numerically. 

Monte Carlo is often the method of choice to evaluate thermodynamic quantities. When the system is classical, system configurations $x$ can be generated randomly according to the thermal distribution $\Pr(x) = \frac{e^{-\beta H(x)}}{\cZ(\beta)}$ by a suitable Markov chain. Sampling from this distribution provides an estimate of the partition function and other quantities of interest.  The running time of the algorithm is determined by the rate of convergence of the Markov chain and the number of samples needed to lower the statistical fluctuations below a desired accuracy. The method of ref.~\cite{SBBK08a} makes it possible to prepare a purification of the thermal distribution in time that scales like the inverse of the square root of the spectral gap.  Building on this work, one of us has recently shown that a quantum computer also achieves a quadratic speed-up with respect to relative accuracy for the task of evaluating partition functions \cite{WCAN08a}. 

Monte Carlo can also be employed for quantum systems using standard mapping between $N$-dimensional quantum systems to $N+1$-dimensional classical systems (the $N+1$th dimension corresponding to inverse temperature). However, frustrated and fermionic systems are affected by the sign problem \cite{MTS87a} which leads to an exponential blow-up of the statistical fluctuations of the sample. For this reason, Monte Carlo simulations are unreliable for these systems. 

This Letter presents two complementary results. First, we demonstrate a quantum algorithm to prepare the thermal Gibbs state $\rho(\beta) = \frac{e^{-\beta H}}{\cZ(\beta)}$ of any locally interacting quantum system, where  $\cZ(\beta) = \Tr (e^{-\beta H})$ is the partition function. The problem of thermalizing a quantum system with a quantum computer was first raised in \cite{TD00a} where heuristic methods were proposed, but no rigorous upper bounds have been derived (for {\em classical} systems the problem was first raised in \cite{LB97a}). Our results set a universal upper bound $D^\alpha$ for the thermalization time of a quantum system, where $D$ is the system's Hilbert space dimension and $\alpha \leq \frac 12$ is proportional to the Helmholtz free energy density of the system.

Second, we present a quantum algorithm to evaluate the partition function of quantum systems within {\em relative} accuracy $\epsilon$. The complexity of the algorithm scales like the inverse of $\epsilon^2$ ($\epsilon$ when the system is classical) and proportionally to the thermalization time of the system. This algorithm is not affected by the sign problem and works for any locally interacting quantum system. It is a full quantum generalization of the algorithm presented in \cite{SBBK08a,WCAN08a} for the evaluation of partitions functions of classical systems. The combination of our two algorithms yields a universal upper bound for the time needed to evaluate a partition function with a quantum computer.

We consider a system composed of $n$ interacting $d$-level particles with a $k$-{\em local} Hamiltonian 
$H = \sum_{j} h_j$ where each term $h_j$ has bounded norm and acts on at most $k$ particles ($k$ is a constant). The most physically relevant case is of course $k=2$.  It is one of quantum information science's most celebrated result that $k$-local Hamiltonians can be simulated efficiently on a quantum computer. More precisely, for those Hamiltonians, the time evolution operator $U(t) = e^{-iHt}$ can be approximately expressed as the product of a sequence of discrete one- and two-qubit gates, and the number of gates scales  essentially linearly with the duration $t$ of the simulated process and polynomially with the system size $n$ \cite{LLo96b, Zal98a, AT03b, BACS07a}.

Given the system's Hamiltonian and a temperature $1/\beta$, we call a {\em thermalization process} any (possibly time-dependent) local Hamiltonian on the system and a bath (of size scaling polynomially with the number of particles in the system) that generates, after time $T$, a unitary transformation $V$ such that
\begin{equation}
\Tr_{\rm bath} \left\{V \left(\frac I D \otimes \kb 00_{\rm bath}\right) V^\dagger\right\} = \rho(\beta),
\label{eq:thermalization}
\end{equation}
or more generally that this equation holds within some accuracy $\delta$. The system's thermalization time $T_{\rm th}(\beta)$ is the duration of the shortest thermalization process. Note that in our definition, the temperature dependence is encoded in the details of $V$ instead of in the initial state of the bath as one would expect physically. This is to rule out trivial thermalization processes where the bath is prepared in the desired state $\rho(\beta)$ and $V$ simply swaps the system and the bath. Nevertheless, the temperature dependence of the thermalization process we will present could equally well be encoded in the initial state of the bath with a simple modification to our algorithm. 

Thermalization processes that occur in nature constitute of local interactions (all forces of nature are two-body), so can be efficiently simulated by a quantum computer, i.e., in a time proportional to the duration of the thermalization process \cite{LLo96b, Zal98a, AT03b, BACS07a}. In what follows, we present a universal thermalization process and characterize its complexity. Then, we show how to use this--or any other--thermalization process to evaluate partition functions  with a quantum computer.

For sake of analysis, we shift the system's Hamiltonian by a constant $E^*$ to ensure that $H$ is positive. This changes the partition function by a factor $e^{-\beta E^*}$ that we keep track of implicitly: we henceforth assume $0<H<E_{\rm max}$ where $E_{\rm max}$ is a known upper bound of the Hamiltonian (since $H$ is $k$-local, $E_{\rm max}$ scales polynomially with the number of particles and can be straightforwardly upper bounded by triangle inequality). We denote the eigenvalues and eigenvectors of the Hamiltonian $H\ket a = E_a\ket a$.

We will make extensive use of the system's time evolution operator with $t = \pi/(4E_{\rm{max}})$, that we simply denote $U$. As mentioned above,  $U$ cannot be simulated exactly with a quantum computer. Instead, we can produce $\tilde U$ that is a good approximation to it. This will unavoidably limit the accuracy of our thermalization process and of the estimated partition function. Indeed, defining the effective Hamiltonian $\tilde H = -\frac it \Log \tilde U$ (by definition of $t$, there is no multi-value problem), our algorithms will prepare the Gibbs state $\tilde\rho(\beta)$ and evaluate the partition function $\tilde \cZ(\beta)$ associated with $\tilde H$, not $H$. However, we show in Appendix A 
 that an accurate simulation $\|U-\tilde U\| \le \epsilon$ leads to an accurate effective Hamiltonian $\| H-\tilde H\| \leq K E_{\rm max}\epsilon$ (where $K$ is a small constant). This, in turn, implies an accurate estimate of the partition function $|\cZ(\beta)-\tilde\cZ(\beta)| \leq 2K\beta E_{\rm max}\epsilon \cZ(\beta)$ (Appendix B 
) and a high fidelity Gibbs state $F[\rho(\beta),\tilde\rho(\beta)] \ge 1-K'\beta E_{\rm max}\epsilon$ (Appendix C 
). These bounds can be understood intuitively by taking a first order expansion in the small $\beta$ limit.

Because the exact time required to simulate $U$ within accuracy $\epsilon$ depends on the details of the system, we measure the complexity of our algorithms in terms of the number of times they need to implement $U$. Hence, the true running time of our algorithms have an additional arbitrarily weak dependence \cite{BACS07a} on the targeted accuracy $\epsilon$ of the partition function and Gibbs state.

Our algorithm requires a cooling schedule  $0=\beta_0 < \beta_1 < \ldots < \beta_\ell=\beta$ such that $F_k := \frac{\cZ(\beta_{k+1})}{\cZ(\beta_{k})}\geq 1/2$ for all $k$. Writing $F_k = \sum_a \frac{e^{-\beta_{k}E_a}}{\cZ(\beta_{k})}e^{-\Delta\beta_k E_a} = \langle e^{-\Delta\beta_k H}\rangle_{\beta_k}$ where $\Delta\beta_k = \beta_{k+1}-\beta_{k}$, we see that a polynomial-length cooling schedule can always be constructed by choosing $\Delta\beta_k =  \log(2)/E_{\rm max}$, which implies a cooling schedule of length $\ell = E_{\rm max}\beta/\log 2$, but shorter schedules can be used in most cases. Notice how the choice of a lower bound on the Hamiltonian  $E^*$ affects these ratios. For this reason, it is desirable that $-E^*$ be as close as possible to the true ground state energy of the system to decrease the length of the cooling schedule. 

The partition function is expressed as a product
\begin{equation}
\cZ(\beta_k) = \cZ(\beta_0) F_0F_1\ldots F_{k-1}
\end{equation}
and will be computed by evaluating each fraction and using the fact that $\cZ(\beta_0) = D$, the dimension of the system's Hilbert space. If each ratio $F_k$ is evaluated within accuracy $\frac\epsilon\ell$, the resulting relative error on $\cZ(\beta)$ will be $\cO(\epsilon)$. Our universal thermalization process requires evaluating the $F_k$ sequentially: the fraction $F_k$ is evaluated from the Gibbs state $\rho(\beta_k)$, and the value of $\cZ(\beta_k)$ is needed to prepare $\rho(\beta_{k+1})$.

Our method makes use of quantum phase estimation (QPE) \cite{Kit95a, NC00a}. This is a transformation that operates on the ``system" register and an $m$-qubit ``energy'' register  initialized to the state $\ket{0_m}$  as follows:
$$\includegraphics{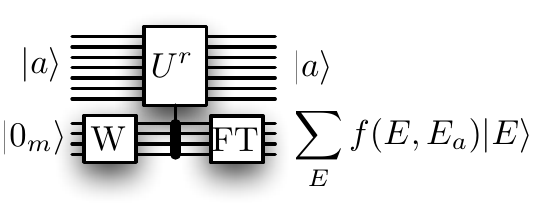} $$
Here, W denotes the Hadamard transform $W\ket{0_m} = 2^{-\frac m2} \sum_{r=0}^{2^m-1} \ket r$ and FT denotes the Fourier transform. The central gate applies $r$ repetitions of the time evolution operator $U$ to the system register, where $r$ is dictated by the state of the lower register. Since $r$ can be as large as $2^m-1$, the running time of this procedure is $2^m$. The label $E$ runs over the discretization of the interval $[0,8E_{\rm max})$, i.e. $E \in \frac{8E_{\rm max}}{2^m}\times \{0,1,\ldots,2^m-1\}$. The function $f(E,E_a)$ is the discrete Fourier transform of $g(r) = e^{-itE_a r}$, so it is highly peaked around the value  $E=E_a$, i.e. 
$$|f(E,E_a)|^2 = \frac{\sin(\frac{E-E_a}{E_{\rm max}}\frac\pi4 2^m)}{\sin(\frac{E-E_a}{E_{\rm max}}\frac\pi4)} .$$ 
Thus, QPE essentially measures the energy of the system and writes it down on the energy register, up to some fluctuations associated to the width of the function $f(E,E_a)$.

Let us first describe our algorithm to prepare Gibbs states by ignoring these fluctuations and return to them later. In fact, our algorithm will prepare {\em purified marked Gibbs states}
\begin{equation}
\ket{\beta_k} = \sum_a \sqrt{\frac{e^{-\beta_k E_a}}{\cZ(\beta_k)}} \ket{a}\otimes\ket{\phi_a} \otimes \ket{E_a}
\end{equation}
that are defined on one system register, one scratchpad register, and one energy register. The term ``purified" refers to the fact that the state $\ket{\beta_k}$ is pure and has a reduced system density matrix equal to the Gibbs state at inverse temperature $\beta_k$. The term ``marked" refers to the fact that the energy register encodes the energy of each system eigenstate.  

To obtain the state $\ket{\beta_0}$, we must first prepare the system and scratchpad in any maximally entangled state $D^{-\frac 12}\sum_a \ket a\otimes \ket{\phi_a}$ and then apply QPE to the system register (ignoring fluctuations of $f$). The finite temperature quantum Gibbs state $\ket{\beta_k}$ is obtained from $\ket{\beta_0} $ and an ancillary qubit initially in the state $\ket 0$ that we rotated by an angle $\theta(E,\beta_k) = \arcsin(e^{-\frac{\beta_k E}{2}})$ conditioned on the energy register, resulting in
\begin{eqnarray}
\ket{\Psi_k} &=&  D^{-\frac 12} \sum_a \ket a \otimes \ket{\phi_a}\otimes  \ket{E_a} \otimes \ket{\theta(E_a,\beta_k)} \\
&=& \sqrt{\frac{\cZ(\beta_k)}{D}} \ket{\beta_k}\ket 0 + \ldots \label{eq:step}
\end{eqnarray}
where $\ket\theta = \sin\theta\ket 0 + \cos\theta\ket 1$ and the ellipsis represents terms in which the ancilla qubit is in the state $\ket 1$. Note that the required conditional rotation can be expressed as the product of $m$ rotations of angle $\arcsin(e^{-\frac{\beta_k E_a}{2^m}})$ controlled by a single qubit. 

We can use Grover's algorithm \cite{Gro96a} to amplify the overlap of $\ket{\Psi_k}$ with the projector $\Pi_0 = I\otimes \kb{0}{0}$ associated to the subspace where the ancillary qubit is in the state $\ket 0$. Indeed, applying $q = \left\lfloor \sqrt{\frac{D}{\cZ(\beta_k)}}\right\rfloor$ times the sequence of two reflections $(I-2\kb{\Psi_k}{\Psi_k})(1-2\Pi_0)$ to the state $\ket{\Psi_k}$ will boost the amplitude of the term $\ket{\beta_k} \otimes \ket{0}$ from its initial value $\sqrt{\frac{\cZ(\beta_k)}{D}}$ to nearly 1. However, $q$ cannot be computed exactly because at this stage of the algorithm, $\cZ(\beta_k)$ is known only within relative accuracy $\frac{k\epsilon}\ell$. This is nonetheless sufficient to amplify the amplitude of the term $\ket{\beta_k} \otimes \ket{0}$ to $\cO(1)$. An amplitude $1-\cO(\frac \epsilon \ell)$ can be achieved using fixed-point search \cite{G05a,H05c} at an additional multiplicative cost of $\log\frac \ell\epsilon$.

The fluctuations of $f(E,E_a)$ will in general invalidate the procedure we have described.  If we perform QPE with a slightly larger number of qubits---i.e., if we use an $(m+4)$-qubit energy register---the probability that the estimated energy $E$ deviates from its true value $E_a$ by more that $2^{-m}/t$ will be less than $1/16$ \cite{NC00a}. To further suppress these fluctuations, we can perform $\eta$ independent QPE procedures and compute the median $M$ of the $\eta$ results. The probability that this median value deviates from the true energy by more than $2^{-m}/t$ is less than  $2^{-\eta}$ \cite{NWZ09a}.

The evaluation of the median can be done by a coherent quantum process. In other words, there is an efficient quantum circuit on $\eta$ energy registers and one $m$-qubit ``median" register that maps $\ket{E^{(1)}}\otimes \ket{E^{(2)}} \otimes\ldots\otimes \ket{E^{(\eta)}} \otimes \ket{0_m}$ to $\ket{E^{(1)}}\otimes \ket{E^{(2)}} \otimes\ldots\otimes \ket{E^{(\eta)}} \otimes \ket{M(E^{(1)},E^{(1)},\ldots)}$.  From the result stated above, we know that on input eigenstate $\ket a$, only two median register states $\ket{E_a^-}$ and $\ket{E_a^+}$ encoding the closest $m$-bit estimates of $E_a$ can have an amplitude of magnitude larger than $\sqrt{2^{-\eta}}$. This can be used to prepare a ``good enough" version of the infinite temperature purified marked Gibbs state
\begin{equation}
\ket{\tilde \beta_0} = D^{-\frac 12} \sum_a \ket a \otimes (\alpha_a \ket{\phi_a^-}\ket{E_a^-} + \beta_a \ket{\phi_a^+}\ket{E_a^+} ) + \ket{\rm bad}
\label{eq:bad}
\end{equation}
where $|\alpha_a|^2 + |\beta_a|^2 \geq 1-2^{-\eta}$,  $|E_a^\pm-E_a| \leq 2^{-m}/t$, and $\ket{\rm bad}$ represent the components where the median deviates from the true energy by more than $2^{-m}/t$, so $\bracket{\rm bad}{\rm bad} \leq 2^{-\eta}$. The scratchpad states $\ket{\phi_a^\pm}$ contain the original states $\ket{\phi_a}$ used to purify the infinite temperature Gibbs state and the $\eta$ (possibly entangled) energy registers used for the computation of the median, whose value is encoded in the third register. 

With $m = \log_2(\beta\ell/t\epsilon)$ the error caused on the Gibbs state by the roundoff of the energy estimate will be $\cO(\frac\epsilon\ell)$ (Appendix~C
).  Similarly, this roundoff error will lead to a $\cO(\frac\epsilon\ell)$ relative error on our estimate of $F_k$ (see Appendix~B
). The error due to the $\ket{\rm bad}$ component of Eq.~\eqref{eq:bad} requires a bit more attention. The reason is that the relative weight of the good and the bad component [i.e., first and second term of Eq.~\eqref{eq:bad}] is in general not preserved by Grover's amplification. Although the norm of $\ket{\rm bad}$ starts out small, it can increase by a factor $e^{\beta_k E_{\rm max}/2}$ during the amplification process. Setting $\eta = [\ln(\frac \ell \epsilon) + \beta_k E_{\rm max}]/\ln(2)$ insures that the norm of $\ket{\rm bad}$ will remain bounded by $\frac\epsilon\ell$ after the amplification process, so can safely be ignored. This completes our presentation of the universal thermalization process, and demonstrates that the Gibbs state can  be prepared with accuracy $\delta = \frac\epsilon\ell$ by a local process in time $T_{\rm th}(\beta) \in \cO\left( \sqrt{\frac{D}{\cZ(\beta)}} \frac{\beta E_{\rm max}}{\delta}\log \frac 1\delta [\log\frac 1\delta +\beta E_{\rm max}]\right)$.

We now describe how the state $\ket{\beta_k}$ can be used to estimate the ratio $F_k$. Similarly to the procedure that led to Eq.~\eqref{eq:step},  we append to the quantum Gibbs state $\ket{\beta_k}$ yet another ancillary qubit in the state $\ket 0$ and rotate it by an angle $\theta(E,\Delta\beta_{k+1})$ conditioned on the energy register, resulting in
\begin{eqnarray}
\ket{\Phi_k} &=&  \sum_a \sqrt{\frac{e^{-\beta_k E_a}}{\cZ(\beta_k)}}\ket a \otimes \ket{\phi_a}\otimes  \ket{E_a} \otimes \ket{\theta(E_a,\Delta\beta_{k+1})} \nonumber \\
&=&  \sqrt{F_k} \ket{\beta_{k+1}}\ket 0 + \ldots \label{eq:step2}
\end{eqnarray}
where the ellipsis represent terms in which the ancilla qubit is in the state $\ket 1$. Clearly, the squared norm of $\Pi_0\ket{\Phi_k}$ is equal to the quantity $F_k$ we are trying to estimate. Thus, $F_k$ can be estimated by quantum counting \cite{BHT98b}, i.e. using phase estimation to estimate the eigenvalue of the unitary matrix $(I-2\kb{\Phi_k}{\Phi_k})(1-2\Pi_0)$ associated to the Jordan block that supports $\ket{\Phi_k}$. An accuracy $\frac\epsilon\ell$ requires $\frac \ell\epsilon$ uses of the circuit that prepares $\ket{\Phi_k}$, so the total time required to evaluate the partition function with accuracy $\epsilon$ with constant success probability is $\cO\left( \sqrt{\frac{D}{\cZ(\beta)}} \frac{\beta^5 E^5_{\rm max}}{\epsilon^2}\right)$.

More generally, any thermalization process can be used to evaluate $F_k$. Indeed, a purified Gibbs state can be prepared by simulating the thermalization process with a quantum computer, substituting the system's maximally mixed state $\frac ID$ in Eq.~\eqref{eq:thermalization} by a maximally entangled state $\frac 1{\sqrt D} \sum_a \ket a\otimes \ket{\phi_a}$. Hence, this requires a time proportional to the thermalization time of the system. This state can be marked using multiple QPEs and evaluating their median as described above. An error at most $\epsilon$ is achieved with constant probability in  time $\cO\left( T_{\rm th}(\beta)\frac{\beta^5 E_{\rm max}^5}{\epsilon ^2}\right)$.

\begin{figure}
\includegraphics[width=7cm]{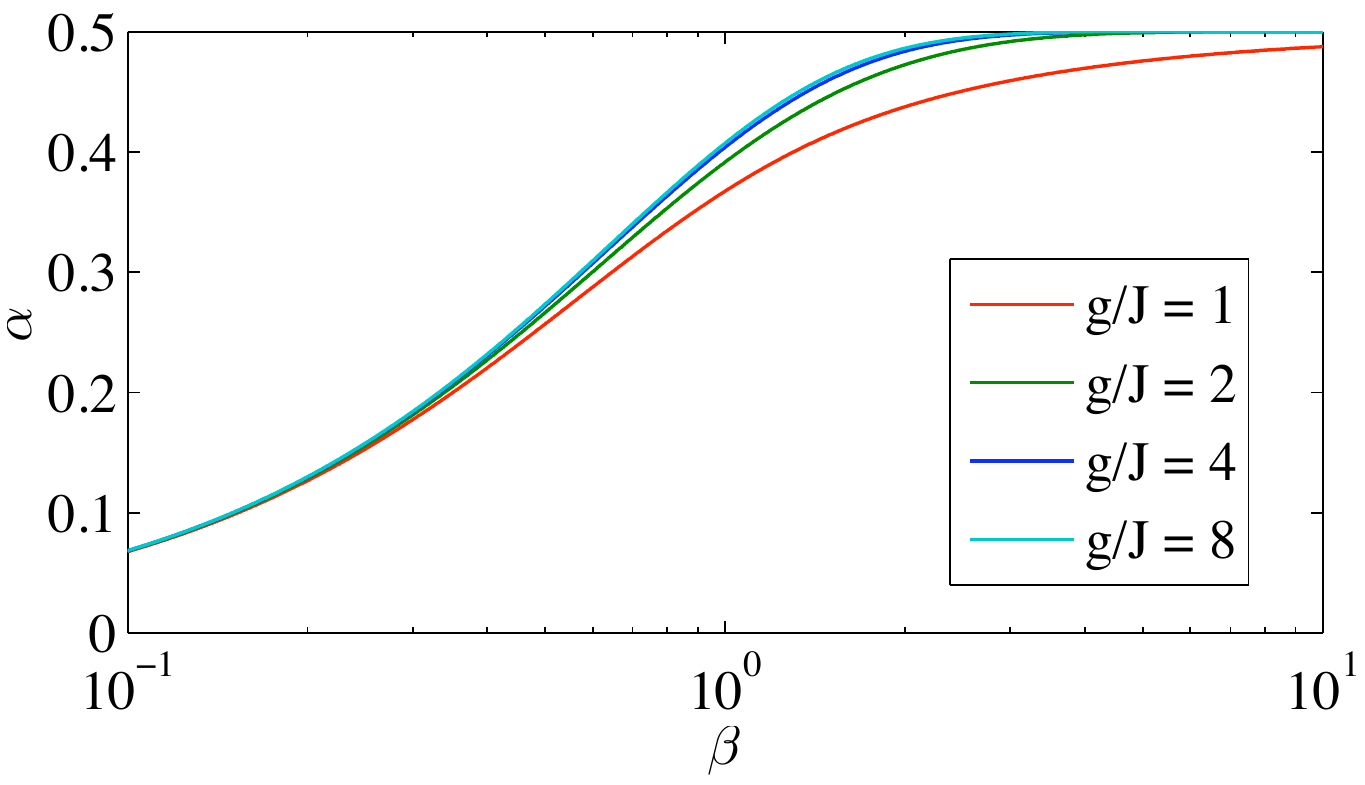}
\caption{(Color online) Scaling exponent as a function of the inverse temperature for the one dimensional Ising model defined by the Hamiltonian $H(J,g) = \sum_i g\sigma_x^{(i)} + J\sigma_z^{(i)}\sigma_z^{(i+1)}$ for different values of $g/J$ (g/J=1 is critical). The absolute values of $g$ and $J$ are fixed by $\|H(g,J)\| = \|H(0,1)\|$. Like the partition function, $\alpha$ is symmetric under $J\leftrightarrow g$.} 
\label{fig:Ising}
\end{figure}

The complexity of our algorithm scales with the system size as $\sqrt\frac{D}{\cZ(\beta)} = D^\alpha$ where $D = d^n$ is the Hilbert space dimension and we can express the scaling exponent $\alpha = \frac 12 h(\beta)/\log(d)$ as a function of the Helmholtz free energy density $h(\beta)$ (energy is defined relative $E^*$). Figure \ref{fig:Ising} shows this exponent for the Ising model as a function of inverse temperature. For temperatures of order unity in natural units, this scaling parameter approaches $\frac 12$, which naturally coincides with the running time found in \cite{PW09a} to compute the ground state energy of the system. At higher temperatures however, the algorithm can be significantly faster. 
Surprisingly, the running time is shortest at the critical field $g/J=1$, where the correlations in the system are the strongest. 

For a classical Hamiltonian, the energy of a given configuration can be computed exactly. In that case, our thermalization process produces a Gibbs state with accuracy $\delta$ in time $\sqrt\frac{D}{\cZ(\beta_k)}\log\frac 1\delta$. The only approximation to the partition function involved in that case comes from quantum counting, so $\cZ(\beta)$ can be evaluated with accuracy $\epsilon$ in time $\cO\left(\sum_k \frac \ell\epsilon \sqrt\frac{D}{\cZ(\beta_k)}\right)$. This should be compared to the time $\cO\left(\sum_k \frac \ell\epsilon \frac 1{\sqrt{\delta_k}}\right)$ required by the quantum simulated annealing algorithm \cite{SBBK08a,WCAN08a}, where $\delta_k$ is the spectral gap of a Markov chain whose fixed point is the Gibbs state of the system at inverse temperature $\beta_k$. Thus, our method offers a speed-up for the computation of a classical partition function whenever no such Markov chain can be found with $\delta \geq \cZ(\beta)/D$. 

\noindent{\em Conclusion---}We have presented a  quantum algorithm to prepare the Gibbs state of a quantum system in a time that grows like the $\alpha$th power of the system's Hilbert space dimension, where $\alpha \leq\frac 12$ is proportional to the  Helmholtz free energy density $h(\beta)$ at inverse temperature $\beta$. This sets a universal upper bound on the thermalization time of quantum systems. However, our universal thermalization process fails to recognize special properties of the system--e.g. large energy gap, absence of long range correlations--that could potentially speed up the computation. We are actively investigating this problem.

We have demonstrated how the ability to thermalize a quantum system on a quantum computer leads to an algorithm to estimate its partition function with accuracy $\epsilon$ in a time  proportional to the system's thermalization time and $\epsilon^{-2}$. Our method provides a complete quantum generalization of the work reported in \cite{SBBK08a,WCAN08a} for partitions functions of classical systems and may provide an additional speed-up when there exist no rapidly mixing Markov chain to prepare the system's Gibbs state.

\bibliographystyle{/Users/dpoulin/archive/hsiam}
\bibliographystyle{/Users/dpoulin/archive/qubib}



\appendix

\section{Simulation error for the effective Hamiltonian}
\label{app:log}
Let $H$ be a self-adjoint operator with $\|H\|\le \pi/4t$ and let $U=e^{it H}$.  Denote by $|a\>$ and $\lambda_a$ the eigenvalues and eigenvectors of $H$, respectively.
Using techniques for simulating Hamiltonian time evolutions, we can realize a unitary $\tilde{U}$ with $\|U-\tilde{U}\| \le \epsilon$.  We prove that there is a self-adjoint operator $\tilde{H}$ such that $\|H-\tilde{H}\|\le \frac{\kappa\epsilon}{t}$ and $e^{it\tilde{H}}=\tilde{U}$ where $\kappa$ is some small constant.

\begin{figure}
\includegraphics[scale=0.83]{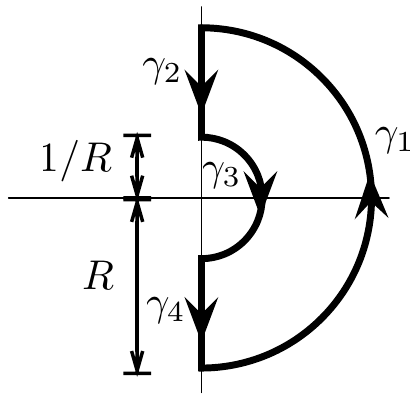}
\caption{For $R>1$, $\gamma$ is the contour defined as the join of the outer arc $\gamma_1(s)=R e^{is}$ for $s\in[\pi/2,\pi/2]$, the upper line segment $\gamma_2(s)=(R-s)i$ for $s\in[0,R-1/R]$, the inner arc $\gamma_3(s)=(1/R) e^{-is}$ for $s\in [-\pi/2,\pi/2]$, and the lower line segment $\gamma_4(s)=(-1/R-s)i$ for $s\in [0,R-1/R]$. We denote by $\gamma^*$, $\gamma_1^*$, $\gamma_2^*$, $\gamma_3^*$, and $\gamma_4^*$ the corresponding images. The contour $\gamma$ is traversed in the positive sense as indicated by the arrows.  } 
\label{fig:contour}
\end{figure}

Let $\C_\pi := \C \setminus (-\infty,0]$.  We use the logarithm
\begin{equation}
\log z = \log|z| + i \arg z\,,
\end{equation}
on $\C_\pi$, where $\arg z$ is defined to be the unique number $\theta\in(-\pi,\pi)$ with $z=|z|e^{i\theta}$. Let $\gamma$ be the contour defined in Fig.~\ref{fig:contour} and denote its interior ${\bf I}(\gamma)$.  

The spectra of $U$ and $\tilde{U}$ are contained in ${\bf I}(\gamma)$.  This is clear for $U$ since $\|tH\|\le \pi/4$. Assume to the contrary that $\tilde{U}$ has an eigenvalue $e^{i\tilde{\lambda}}$ with $|\tilde{\lambda}|\ge \pi/3$.  Let $|\psi\>$ denote the corresponding eigenvector of $\tilde{U}$ and $p_a=|\<a|\psi\>|^2$.  For $\epsilon \leq 1/4$, this leads to the following contradiction
\begin{eqnarray}
\epsilon 
& \ge &
|\<\psi |U|\psi\> - \<\psi|\tilde{U}|\psi\>| \\
& = &
\big| \sum_a p_a \big( e^{i\lambda_a} - e^{i\tilde{\lambda}} \big) \big| \\
& \ge &
|e^{i\frac{\pi}{4}} - e^{i\frac{\pi}{3}}|= 2\sin\frac\pi{24} \approx 0.26 > \frac 14\,.
\end{eqnarray}
Generalizing this argument and using the inequality $\phi \leq 2\sin\phi$ for $\phi \in [0,\pi/2]$, we arrive at the inequality $\|\tilde H + c\| \leq \|H+c\| + \epsilon/t$ for all real $c$. This implies that the spectrum $\sigma(\tilde H)$ of $\tilde H$ is contained in the interval $(\min\{\sigma(H)\}-\frac\epsilon t,\max\{\sigma(H)\}+\frac\epsilon t)$. This result will be used in Appendix~\ref{app:rho}. 

For our purposes, it is important that the logarithm function defined above is Lipschitz continuous in ${\bf I}(\gamma)$.  More precisely, we have
\begin{equation}\label{eq:lipschitz}
|\log a - \log b | \le \kappa |a - b| \,.
\end{equation}
This is seen as follows.  We may assume w.l.o.g. that $a=|a|$ and $b=|b| e^{i\theta}$ with $\theta\in[0,\pi]$.  We have
\begin{equation}
| \log a - \log b |^2 = (\log|a| - \log |b|)^2 + \theta^2
\end{equation}
and 
\begin{equation}
|a - b|^2 = (|a| - |b|)^2 + |a||b|4\sin^2\frac\theta 2\,.
\end{equation}
Using the fact that $1/R \leq |a|,|b| \leq R$ in ${\bf I}(\gamma)$, we clearly have 
\begin{equation}
| \log a - \log b |^2 \le R^2 (|a| - |b|)^2
\end{equation} 
since the real logarithm is Lipschitz continuous with constant $R$ on the interval $[1/R,\infty)$, and 
\begin{equation}
\theta^2 \leq \frac{4}{R^2} |a||b| 4\sin^2\frac\theta 2 
\end{equation}
following from the inequality $\phi \leq 2\sin\phi$ on $[0,\frac\pi 2]$. Thus, the Lipschitz constant is $\kappa = \max\{\frac 2 R, R\}$.

Let $A$ be an arbitrary diagonalizable matrix and $S$ an invertible matrix such that 
\begin{equation}\
A=S \, \mathrm{diag}(a_1,\ldots,a_n) \, S^{-1}\,.
\end{equation}
Assume that $a_1,a_2,\ldots,a_n$ are in $\C_\pi$.  Then, it is natural to define the logarithm of $A$ as follows
\begin{equation}\label{eq:logDiag}
\Log A = S \, \mathrm{diag}(\log a_1, \log a_2, \ldots, \log a_n) \, S^{-1}\,.
\end{equation}
It is possible to extend the definition of the logarithm to all matrices $A\in M_n(\C)$ with spectrum $\sigma(A)\subset \C_\pi$ with the help of the Jordan canonical form of $A$.  This gives rise to the so-called primary matrix function $\Log A$ associated to the scalar-valued stem function $\log z$.  We refer the reader to \cite[Chapter 6 ``Matrix Functions'']{HJ91a} for more details.

Since $U$ and $\tilde{U}$ are unitary, and thus diagonalizable, we can define their logarithms using the formula in eq.~(\ref{eq:logDiag}). It is clear that 
\[
\Log U = itH\,.
\]
We would like to be able to conclude that $\| \Log U - \Log \tilde{U}\|$ is small since $\|U-\tilde{U}\|\le\epsilon$ is small.  This is true when $U$ and $\tilde{U}$ commute.  In this case, we may assume that they are both diagonal.  Hence, everything reduces to eq.~(\ref{eq:lipschitz}) since it suffices to compare corresponding diagonal entries.

To prove the desired result in the non-commutative case, we need to consider the following result which is established in Theorem~6.4.20 in \cite{HJ91a}. Let ${\cal D}=\{ A \in M_n(\C) \, : \, \sigma(A) \subset {\bf I}(\gamma)\}$.  For any $A\in{\cal D}$, define
\begin{equation}
\Log A = \frac{1}{2\pi i}\oint_\gamma (\log z) (zI - A)^{-1} dz\,.
\end{equation}
Then, $\Log$ is continuous on ${\cal D}$, and
\begin{equation}
e^{\Log A} = A\,.
\end{equation}

Let $A$ and $B$ be two arbitrary {\em unitary} matrices with $\sigma(A), \sigma(B) \subset {\bf I}(\gamma)$.  We now show that
\begin{equation}
\| \Log A - \Log B \| \le K \| A - B \|
\end{equation}
holds for all such unitaries $A$ and $B$, where $K$ is some constant.  We have
\begin{align}
&\| \Log A - \Log B \| \\
& =
\| \frac{1}{2\pi i} \oint_\gamma (\log z) [(zI - A)^{-1} - (zI - B)^{-1}] dz \| \\
& =  
\frac{1}{i2\pi} \oint_\gamma  \log z  \, (zI - A)^{-1} (A - B)  (zI - B)^{-1}  \, dz \\
& \le  
\frac{1}{2\pi} \oint_\gamma | \log z | \, \| (zI - A)^{-1} \| \, \| A - B \| \, \| (zI - B)^{-1} \| \, |dz|\,.  \label{eq:last}
\end{align}
We need to bound $|\log z|$, $\| (zI - A)^{-1} \|$, and $\| (zI - B)^{-1} \|$ on the outer and inner arcs $\gamma_1^*$ and $\gamma_3^*$, and the upper and lower line segments $\gamma_2^*$ and $\gamma_4^*$.

We have $|\log z| = \big| \log |z| + i \, \mathrm{arg}(z) \big| \le \sqrt{\log^2(R) + \pi^2}$ on $\gamma^*$.  On $\gamma_1^*$ we have
\begin{eqnarray}
\| (z I - A)^{-1} \| 
& = &
|z^{-1}| \, \| (I - z^{-1} A)^{-1} \| \\
& = &
|R^{-1}| \, \| \sum_{k=0}^{\infty} (z^{-1} A)^k \| \\
& \le &
|R^{-1}| \, \sum_{k=0}^{\infty} \| (z^{-1} A) \|^k \\
& \le &
|z^{-1}| \, | (1- \| z^{-1} A \|)^{-1} \\
& = & (R-1)^{-1}\,.
\end{eqnarray}
We used the fact that the above power series converges since $\| z^{-1} A \| = 1/R < 1$.  

We cannot use the same argument on the $\gamma_3^*$ because $\| z^{-1} A\| = R$ and the power series would diverge.  Therefore, we rewrite $(z I - A)^{-1}$ as
\begin{equation}
(z I - A)^{-1} = (z A A^{-1} - A)^{-1} = - A^{-1} (I - z A^{-1})^{-1}\,.
\end{equation}
The norm of $A^{-1}$ is $1$, so it suffices to show that the norm of $(I - z A^{-1})^{-1}$ bounded.  To do this, we can use the power series expansion since $\|z A^{-1} \|=1/R$ on $\gamma_3^*$.  We obtain
\begin{equation}
\| (I - z A^{-1})^{-1} \| \le \frac{R}{R-1}\,.
\end{equation}
On $\gamma_2^*$ and $\gamma_4^*$ we have $\|(zI - A)^{-1}\| \le 2$ since $\pm e^{i\frac{\pi}{3}}$ are the closest points to the imaginary axis, which could be in $\sigma(A)$. The distance $e^{i\frac{\pi}{3}}$ and the imaginary axis is $\mathrm{Re}\, e^{i\frac{\pi}{3}} = 1/2$.  The same bounds are true for $(zI - B)$.

The length $|\gamma|$ of $\gamma$ is $\pi(R + 1/R) + 2(R-1/R)$.  Finally, we obtain the desired Lipschitz constant $K$ by choosing $R=2$ which yields $K \le 6\pi+12$.

\section{Lipschitz continuity of partition functions with respect to Hamiltonians}
\label{app:Z}

Let $\tilde{H}$ and $H$ be two self-adjoint operators with $\tilde{H}=H + E$ and $\| E \| \le \frac\epsilon t$.  Denote by 
${\cZ}(\beta)$ and $\tilde\cZ(\beta)$ the corresponding partition functions at (inverse) temperature $\beta$. We prove 
that
\begin{equation}
(1/\delta) \cZ(\beta) \le \tilde{\cZ}(\beta) \le \delta \cZ(\beta)
\end{equation}
holds for $\delta=e^{\beta\epsilon/t}$. This makes it possible to estimate $\cZ(\beta)$ with small relative error by estimating $\tilde{\cZ}(\beta)$ instead provided that the norm of the error term $E$ is sufficiently small for given temperature $\beta$.

This readily follows from Weyl's perturbation theorem \cite{Bha96}.  Let $A$ and $B$ be two arbitrary self-adjoint operators with eigenvalues
$\lambda^\downarrow_1(A)\ge \ldots \ge \lambda^\downarrow_n(A)$ and $\lambda^\downarrow_1(B)\ge \ldots \ge \lambda^\downarrow_n(B)$, respectively. Then,
\begin{equation}
\max_j | \lambda^\downarrow_j(A) - \lambda^\downarrow_j(B) | \le \|A - B\|\,.
\end{equation}
This implies 
\begin{equation}
e^{-\|A - B\|} \, \tr e^A \le \tr e^B \le e^{\|A - B\|} \, \tr e^A\,.
\end{equation}
We obtain the desired result by setting $A=-\beta H$ and $B=-\beta\tilde{H}$.

\section{Lipschitz continuity of Gibbs states with respect to Hamiltonians}
\label{app:rho}

Let $\tilde{H}$ and $H$ be two self-adjoint operators with $\tilde{H}=H + E$ and $\| E \| \le \frac\epsilon t$ where $t=\pi/(4 E_{\max})$.
Denote by 
$\rho(\beta) = \exp(-\beta H)/\cZ(\beta)$ and $\tilde\rho(\beta) = \exp(-\beta\tilde H)/\tilde\cZ(\beta)$ the corresponding Gibbs states at (inverse) temperature $\beta$. We prove that the fidelity between $\rho(\beta)$ and $\tilde{\rho}(\beta)$ is high.  More precisely, we have
\begin{equation}
F(\rho(\beta),\tilde\rho(\beta)) \geq e^{-\beta\epsilon/t}\,.
\end{equation}

Recall that the quantum fidelity is defined by
\begin{equation}
F(\rho,\tilde{\rho}) = \tr \sqrt{\sqrt{\rho}\tilde{\rho}\sqrt{\rho}}\,.
\end{equation}
We bound the fildelity $F(\rho,\tilde{\rho})$ from below for the case of Gibbs states $\rho=\exp(-\beta H)/\cZ(\beta)$ and $\tilde{\rho}=\exp(-\beta \tilde{H})/\tilde{\cZ}(\beta)$ by relying upon the following inequality.  It was proved in \cite{FS93} that the function $p \mapsto \tr \big( e^{pB/2} e^{pA} e^{pB/2} \big)^{1/p}$ is increasing for self-adjoint operators $A$ and $B$ in $p\in(0,\infty)$.  Its limit at $p=0$ is $\tr e^{A+B}$.  In particular, the inequalily
\begin{equation}
\tr e^{A+B} \le \tr \Big( e^{pB/2} e^{pA} e^{pB/2} \Big)^{1/p}
\end{equation}
holds for every $p>0$.  This is a strengthened variant of the Golden-Thompson inequality, which is obtained for $p=1$.

Setting $p=2$, $A=-\beta \tilde{H}/2$, and $B=-\beta H /2$, we bound the fidelity from below in terms of partition functions as follows
\begin{eqnarray}
F(\rho,\tilde{\rho}) 
& = &
\frac{ \tr \Big( e^{-\beta H/2} e^{-\beta \tilde{H}} e^{-\beta H/2} \Big)^{1/2}}{\sqrt{\cZ(\beta) \tilde{\cZ}(\beta)}} \\
& \ge &
\frac{ \tr \Big( e^{-\beta(H + \tilde{H})/2} \Big)}{\sqrt{\cZ(\beta) \tilde{\cZ}(\beta)}} \\
& = &
\frac{\hat{\cZ}(\beta)}{\sqrt{\cZ(\beta) \tilde{\cZ}(\beta)}}\,,
\end{eqnarray}
where $\hat{H}=(H+\tilde{H})/2$.  Note that $\|H-\hat{H}\|\le \epsilon/(2t)$.
Using the results of Appendix~B, we obtain
\begin{equation}
F(\rho,\tilde{\rho}) \ge \frac{\hat{\cZ}(\beta)}{\sqrt{\cZ(\beta) \tilde{\cZ}(\beta)}} 
\ge 
e^{-\beta\epsilon/t}\,.
\end{equation}

\end{document}